\documentclass{aa}
\usepackage{graphicx}
\usepackage{natbib}
\usepackage{txfonts}

\begin{document}

\title{Visibility of unstable oscillation modes in a rapidly rotating B star}

\author{G.J. Savonije}

\institute{Astronomical Institute `Anton Pannekoek', University of
  Amsterdam,   Science Park 904 1098 XH Amsterdam, The Netherlands}

\date{Received ; accepted }

\abstract 
{Space missions like CoRoT and Kepler have provided numerous new observations of stellar oscillations in a multitude of stars by high precision photometry.  The identification of the photometrically observed oscillations is, however, difficult and requires detailed model calculations of pulsating stars.} 
{This work compares the observed rich oscillation spectrum of the rapidly rotating  B3 IV star HD 43317 with the first results obtained by a new method to calculate unstable oscillation modes in rapidly rotating stars in order to see whether some of the observed modes can be identified.}
{The new numerical method consists of two parts. We first search for modes resonant with a prescribed forcing symmetry by moving through relevant regions of complex frequency space and monitoring any increase of the stellar response to the applied forcing and zooming in onto the resonance. These resonant non-adiabatic 2D-solutions are then fed into a 2D relaxation code with the same equations but without forcing terms. The complex oscillation frequency used in the forcing is now no longer prescribed, but added as an extra unknown. The corresponding free oscillation mode is usually obtained after a few ($<10$) iterations with only minor adjustment of the complex oscillation frequency. To compare with the observed light variations we calculate the `visibility' of the found unstable oscillation modes, taking into account the cancellation of the various parts of the radiating oscillating stellar surface as seen by the observer. }
{The frequencies of unstable axisymmetric g-modes, which have the highest visibility, appear to nearly coincide with the observed largest amplitude photometric variations of HD 43317, making an identification of the latter oscillations as $m$=0 modes plausible. The identification of $m$=1 g-modes is less straightforward, while many of the unstable even $m$=2 g-modes may correspond to observed weaker photometric variations. Only one unstable r-mode has non-negligible visibility. The observationally inferred almost equidistant period spacings of ten, respectively seven, oscillation frequencies for HD 43317 cannot be reproduced. } 
{}

\maketitle

\section{Introduction}
The Kepler \citep{Bor09} and CoRoT (\citet{Auv09},  \citet{Bag09}) space missions have resulted in a wealth of detailed new pulsation data for an extended sample of stars.
\citet{Papi12} presented a list of oscillation frequencies obtained by 5 months of photometric monitoring of the B3 IV star HD 43317 by the CoRoT satellite  combined with high resolution and high S/N spectra obtained with the ground based HARPS instrument of ESO.  They conclude from the spectroscopic variability that for this star the rotation frequency $\Omega_\mathrm{s}$ is about 50 \% of the critical break-up speed. Asteroseismic studies of rapidly rotating stars are complicated by the fact that the Coriolis force makes the set of oscillation equations unseparable into an $r$ and $\theta$ part. A possible approach \citep{Unno89} is to approximate the full oscillation solution by a truncated series of spherical harmonics to describe the $\mathrm{r}$ and $\theta$ dependence and solve a coupled set of differential equations by a shooting method. One may assume a $\varphi $ dependence of $e^{\mathrm{i} m \varphi}$ with given $\mathrm{m}$, since the unperturbed star is assumed spherically symmetric.  However, many of the observed oscillation frequencies with significant amplitudes in \citet{Papi12} fall in the so-called 'inertial-range' with  $|\overline{\sigma}/(2 \Omega_\mathrm{s})| < 1$, for which the corresponding oscillation modes  cannot be adequately described by superposition of only a small number of spherical harmonics, which makes this method rather cumbersome.

Several papers describing new developments with oscillation codes have appeared recently, see \citet{Ouaz12}, \citet{Vals13} and \citet{Town13}.
In this paper another approach is used whereby an iterative relaxation calculation (like the Henyey codes used for stellar evolution calculations) is performed on a fine ($r$, $\theta$) grid using a  2D non-adiabatic code in which the Coriolis force (first order in $\Omega_s$) is fully taken into account but the centrifugal force (being second order in $\Omega_s$) is neglected. 
We use a two-step procedure where we first search in complex frequency space  for strong stellar responses using a prescribed forcing as done earlier \citep{Sav05}. After zooming in to these resonances we now use in the second step the thus obtained resonant response  as input for the relaxation code (in which the forcing terms have been taken out) which iterates until the solution converges to the corresponding free oscillation mode.
In this way it is relatively easy to find free oscillation modes with a relaxation code. A problem (even with 1D) relaxation codes has always been to find a suitable starting model from which the oscillation code is able to succesfully converge the iterative sequence, e.g. \citet{Unno89}. With the current code this is generally no longer a problem. The relaxation method where the oscillation equations are solved as difference equations on a 2D grid has the advantage that the higher order harmonics induced by the Coriolis force are included in the solution. Evidently it requires a grid with adequate resolution which implies use of considerable computer power.

 Although the inferred stellar parameters \citep{Papi12} imply a rapid rotation speed of $\Omega_\mathrm{s}/\Omega_\mathrm{c}$= 0.5 that would require inclusion of the centrifugal distortion  ($\propto (\Omega_\mathrm{s}/\Omega_\mathrm{c})^2$) of the star, a spherically symmetric star is used as input model for the pulsation calculations.  Note in this respect that the non-spherical centrifugal distortion of the star induces  baroclinic currents in the otherwise stably stratified radiative layers which can significantly perturb the angular momentum profile. This is a significant complication which makes it difficult to produce an unperturbed non-spherical equilibrium structure, but see the work by \citet{Esp13} for recent developments.

 In the current oscillation calculations the Coriolis force is fully taken into account as this force is required to simulate the direct first order effect of rotation on stellar pulsations.
 We study the oscillation modes and their linear stability in a  B-star rotating with $\Omega_s/\Omega_c=0.5$ in the frequency range where the strongest photometric amplitudes are observed in HD 43317 (up to 4.5 cycles per day in the observer's frame).

 \subsection{Stellar input model}
 For a first test of this new code we consider a uniformly rotating main sequence star of 5.4~$\mbox{${\rm M}_{\odot}$}$ with radius $R_\mathrm{s}= 3.8 \ \mbox{${\rm R}_{\odot}$}$,  $T_\mathrm{eff}=1.70 \times 10^4 \ \mathrm{K}$ and Z=0.014 as a model for the B3 IV star HD 43317, in line with data given in \citet{Papi12}.  The unperturbed  (1D) stellar model used as input model for the oscillation calculations was obtained with the freely obtainable (http://mesa.sourceforge.net/) stellar evolution code MESA \citep{Pax13} using standard OPAL opacities and no overshooting from the convective core. The 5.4 $M_{\odot}$ model has a convective core that extends to $r/R_s$~ = ~0.118 and a shallow convective surface shell between $r/R_s = 0.994-0.996$.

\section{Basic oscillation equations}
\label{basics}
 We use spherical coordinates ($r,\theta,\varphi)$ with the origin at the star's centre, whereby $\theta$ = 0  corresponds to its rotation axis. We give here the forced version of the oscillation equations (with the forcing terms $\mathcal{S}$ or $\mathcal{T}$)  used to calculate the input model for the relaxation code (where these terms are set equal to zero).

Let us denote the velocity perturbation and displacement vector in the star by $\vec{v^\prime}$ and  $\vec{\xi}$ with $\vec{v^\prime}={\rm i} \, \sigma \, \vec{\xi}$ and denote perturbed Eulerian quantities like pressure $P'$, density $\rho'$, temperature $T'$, and the energy flux vector $\vec{F'}$ with a prime. The linearized hydrodynamic equations governing the non-adiabatic oscillations of a star in the corotating frame are \citep{Unno89}:
\begin{eqnarray}
\left[ \left(\frac{\partial}{\partial t} + \Omega_\mathrm{s} \frac{\partial }
    {\partial\varphi}\right)  v'_i\right] \vec{e}_i+ 2 \vec{\Omega}_\mathrm{s} \times \vec{v'}
  +\frac{1}{\rho} \nabla P' - \frac{\rho'}{\rho^2} \nabla P = \mathcal{S}_F
   \label{eqmot}\label{eq:1}
\end{eqnarray}
\begin{equation}
  \left(\frac{\partial}{\partial t} + \Omega_\mathrm{s} 
    \frac{\partial }{\partial\varphi}\right) \rho' +
  \nabla\cdot\left(\rho \vec{v'} \right) =0, \label{eqcont}
\end{equation}
\begin{equation} 
  \left(\frac{\partial }{\partial t} + \Omega_\mathrm{s}
    \frac{\partial }{\partial \varphi}\right) \left[ s' +
    \vec{\xi}\cdot \nabla s \right]=-\frac{1}{\rho T} \nabla\cdot\vec{F'}, \label{eqe}
\end{equation}
\begin{equation}
  \frac{\vec{F'}}{F}=\left(\frac{\mathrm{d}T}{\mathrm{d}r}\right)^{-1} \left[ \left(\frac{3
        T'}{T} -\frac{\rho'}{\rho} -\frac{\kappa'}{\kappa} \right) \nabla
    T + \nabla T' \right] \label{eqf} 
\end{equation}
where we have added the forcing term $\mathcal{S}_F$ , while ${\bf e}_i$ are the unit vectors of our spherical coordinate system, $\kappa$ denotes the opacity of stellar material and $s$ its specific entropy. These perturbation equations represent, respectively, conservation of momentum, conservation of mass and
conservation of thermal energy, while the last equation describes the
perturbed radiative energy diffusion. For simplicity we adopt the usual  \citet{Cow41} approximation: i.e. we neglect perturbations to the gravitational potential caused by the star's oscillatory distortion. For the oscillation modes studied here this is an adequate approximation.

The unperturbed stellar model being spherically symmetric, we may choose a fixed value of the azimuthal index $m$ (plus the oscillation frequency  $\sigma$ relative to the inertial frame) and separate the $\varphi$- and time part in the oscillation equations given above. Thus we may write the solution for the displacement vector etc.  as  
\begin{equation}
\vec{\xi}(r,\theta,\varphi,t)\:=\: \vec{\overline{\xi}}(r,\theta)\: \rm{e}^{\rm{i}\left( \sigma t -m \varphi\right) }.
\end{equation}

From now on the displacements and Eulerian perturbations are considered functions of $r$ and $\theta$ only and the above overline on $\xi(r, \theta)$ and the factor ${\rm e}^{\rm{i} (\sigma t - m \varphi)}$ will be omitted in the equations. Note that, in contrast to the usual practice, we consider $m$ to be always positive. Then $\sigma >0$ corresponds to wave motion in the positive $\varphi$ direction and negative $\sigma$  corresponds to retrograde wave motion in the inertial frame.

After defining a certain value for the azimuthal index $m$ the star is forced with a spherical harmonic component  $(l,m)$ whereby the order $l$ is chosen as the smallest integer for the adopted equatorial symmetry (either odd or even). The oscillation frequency in the  corotating frame is $\overline{\sigma} = \sigma - m \mbox{$\Omega_{\rm s}$}$ and negative/positive for retrograde/prograde modes in the frame corotating with the star. Obviously $\overline{\sigma}=0$ corresponds to $\sigma/m=\mbox{$\Omega_{\rm s}$}$, i.e. the observer would see a mode with pattern speed equal to the angular rotation rate of the star.   Assuming a uniformly rotating star with angular velocity $\mbox{$\Omega_{\rm s}$}$, the radial, $\theta$ and $\varphi$ components of the equation of motion can  be expressed as

\begin{eqnarray}
\overline{\sigma}^2 \xi_{r} + \left( 2 \rm{i} \overline{\sigma}\mbox{$\Omega_{\rm s}$}  \sin {\theta} \right)  \xi_\varphi -\frac{1}{\rho} \,\frac{\partial P'}{\partial  r } 
+ \frac{1}{\rho} {{\rm d} P\over {{\rm d} r}} \left( \frac{\rho'}{\rho}\right) = \mathcal{S}_r \label{EQ1}
\end{eqnarray}

\begin{eqnarray}
 \overline{\sigma}^2   \xi_\theta  +  \left( 2  \rm{i}  \overline{\sigma}\mbox{$\Omega_{\rm s}$}  \cos{\theta} \right)  \xi_\varphi 
- \frac{P}{\rho}\, \,\frac{1}{r} \frac{\partial }{\partial \theta }\left( \frac{P'}{P}\right) = \mathcal{S}_\theta
\label{EQ2}
\end{eqnarray}

\begin{eqnarray}
\overline{\sigma}^2  \xi_{\varphi} -  2  \rm{i} \overline{\sigma}\mbox{$\Omega_{\rm s}$}
\left( \sin {\theta} \, \xi_{r} + \cos {\theta} \, \xi_{\theta}\right) 
+\frac{\rm{i} m }{r\sin{\theta}} \frac{P}{\rho} \left( \frac{P'}{P} \right)  = \mathcal{S}_\varphi \label{EQ3}
\end{eqnarray}

The spheroidal forcing terms on the right hand side of equations (\ref{EQ1})-(\ref{EQ3}) are applied in case of a predominantly spheroidal (g- or p-) mode and are defined (using an arbitrary scaling constant $C$) as
\noindent

\[ \mathcal{S}_r =  C \, l \,\frac{r^{l-1}}{\rho} \,\rm{P}^{m}_{l}(\theta)   \]
 \begin{eqnarray}
 \mathcal{S}_\theta = C \: \frac{r^{l-1}}{\rho} \,\frac{\partial }{\partial \theta }\, \rm{P}^{m}_{l}(\theta) \nonumber &&
\: ; \: \mathcal{S}_\varphi =  - {\rm i} \, m \, C \,\frac{r^{l-1}}{\rho} \,\frac{{\rm{P}}^{m}_{l}(\theta)}{\sin{\theta}} \nonumber
\end{eqnarray}
 When a predominately toroidal (r-)mode is studied the horizontal forcing terms $\mathcal{S}_\theta$ and $\mathcal{S}_\varphi$ in equations (\ref{EQ2}) and (\ref{EQ3}) are (adequately) replaced by $\mathcal{T}_\theta$ and $\mathcal{T}_\varphi$ with 
\begin{eqnarray}
\mathcal{T}_\theta = - {\rm{i}} \, m \,C \, \frac{r^{l'-1}}{\rho} \, \frac{{\rm{P}}^{m}_{l^{\prime}}(\theta)}{\sin{\theta}} \nonumber &&
; \: \mathcal{T}_\varphi = - C \, \frac{r^{l'-1}}{\rho} \, \frac{\partial {\rm{P}}^{m}_{l^{\prime}}(\theta)}{\partial \theta }
\nonumber
\end{eqnarray}
where for even (with $l-m$ even) modes $ l^{\prime}=l+1$ and for odd modes (with $l-m$ odd) $l^{\prime}=l-1$.

These forcing terms are applied to search for resonant modes by varying the complex forcing frequency in the direction of maximum response. The found resonant modes are then used as input to obtain the corresponding free oscillation modes.

The equation of continuity (\ref{eqcont}) can be expressed as
\begin{eqnarray}
\frac{\rho^{\prime}}{\rho} + \frac{1}{r^2 \rho} \frac{\partial }{\partial r }
\left( \rho r^2 \, \xi_{r}\right) + \frac{1}{r \sin{\theta}}  \frac{\partial }{\partial \theta } \left( \sin{\theta} \, \xi_\theta\right) - \frac{\rm{i}\, m}{r  \sin{\theta}}  \xi_\varphi \: = \: 0   \label{EQ4}
\end{eqnarray}

After applying the thermodynamic relation 
\[ \delta s = \frac{P}{\rho T}\frac{1}{\Gamma_3-1}
\left(\frac{\delta P}{P} -\Gamma_1 \frac{\delta \rho}{\rho} \right) \]
(where $\delta s$ denotes a Langrangian entropy perturbation) in terms of Chandrasekhar's adiabatic  exponents $\Gamma_1$ and $\Gamma_3$, the energy equation (\ref{eqe}) can be rewritten 

\begin{eqnarray}
\frac{1}{\Gamma_1} \frac{P'}{P} - \frac{\rho^{\prime}}{\rho} - \mathcal{A} \, \xi_{r} = {\rm{i}} \,\Lambda  \left[ \nabla \cdot \left( \frac{\vec{F'}}{F}\right)  + {{{\rm d} \ln F}\over {{\rm d} r}} \left( \frac{F'_r}{F}\right) \right] \label{EQ5}
\end{eqnarray}
where $F$ is the unperturbed (radial) energy flux. The Schwarzschild discriminant $\mathcal{A}$ and characteristic diffusion length $\Lambda$ are defined by:

\begin{eqnarray}\mathcal{A} = \frac{\rm{d} \ln{\rho}}{\rm{d} r }- \frac{1}{\Gamma_1} \frac{\rm{d} \ln{P}}{\rm{d} r } \: ; && \; \Lambda = \frac{(\Gamma_3 -1)}{\Gamma_1} \frac{ {\rm F}}{\overline{\sigma} P} \nonumber \end{eqnarray}

The radial and horizontal components of the perturbed energy flux follow from equation (\ref{eqf}) as
\begin{equation} 
\frac{F^{\prime}_r}{F}  =\left( {{{\rm d} \ln{T}}\over {{\rm d} r}}\right)^{-1}  \frac{\partial }{\partial r  }\left( \frac{T'}{T}\right)  - \left( \kappa_T -4 \right) \left( 
\frac{T'}{T}\right)  - \left( \kappa_{\rho} + 1 \right) \left( \frac{\rho^{\prime}}{\rho} \right) 
\label{EQ6}
\end{equation}

\begin{equation}
\frac{F'_\theta}{F} = \left({{{\rm d} \ln{T}}\over {{\rm d} r}}\right)^{-1} \, \frac{1}{r} 
\frac{\partial }{\partial \theta } \left( \frac{T'}{T} \right) \label{eqFth}
\end{equation}

\begin{equation}
\frac{F'_\varphi}{F} = \frac{- \rm{i} \,m} {r \sin{\theta}} \left({{{\rm d} \ln{T}}\over {{\rm d} r }}\right)^{-1} \left( \frac{T'}{T} \right)  \label{eqFph}
\end{equation}

After substituting $F'_\theta$ and $F'_\varphi$ in the divergence term of equation (\ref{EQ5}) 
and using the equation of state (\ref{EOS}) (where the $\chi$'s stand for the usual logarithmic partial derivatives of the pressure and
$\chi_\mu= \chi_\rho \, {{{\rm d} \ln{\rho}}\over {{\rm d} r}} + \chi_T \, {{{\rm d} \ln{T} }\over {{\rm d} r }}-{{{\rm d}\ln{P}}\over {{\rm d} r}}$
is of importance in the layers outside the convective core with a composition gradient):
\begin{equation}
\frac{P'}{P} = \chi_\rho \,\frac{\rho^{\prime}}{\rho} + \chi_T \, \frac{T'}{T}  + \chi_\mu \,\xi_r
\label{EOS}
\end{equation}
 to eliminate $\rho^\prime$ from equations (\ref{EQ1}) - (\ref{EQ6}) we are left with six equations in the six unknowns $(\xi_r/{\rm{R_s}})$, $(\xi_\theta/{\rm{R_s}})$, $(\xi_\varphi/{\rm{R_s})}$, $(P'/P)$, $(T'/T)$ and $(F'_r/F)$.

In convective regions, i.e. in the convective core and convective shell at the surface of our $5.4 \mbox{${\rm M}_{\odot}$}$ B-star, we add (turbulent) viscous terms 
\begin{equation}
\frac{{\rm i}\, \overline{\sigma}}{\rho\, r^2} \,\frac{\partial }{\partial r } \left(\rho \, \zeta \, r^2\,\frac{\partial \xi_i}{\partial  r}\right) + \frac{{\rm i}\, \zeta \,  \overline{\sigma}}{r^2 \, \sin{\theta}} \,\frac{\partial }{\partial \theta }\left(\sin{\theta} \,\frac{\partial \xi_i}{\partial \theta  } \right)
\end{equation}

 to the equations of motion (\ref{EQ1})-(\ref{EQ3}) where the subindex $i$ stands for $r$,  $\theta$ or $\varphi$ and the kinematic viscosity $\zeta$ is calculated with simple mixing length theory. The mixing length $\lambda= 1.5 \, H_P$ and convective velocities are taken from the stellar evolution code MESA that was used to calculate the unperturbed equilibrium model. The kinematic viscosity $\zeta$ is corrected (reduced) for timescale mismatch between  convective- and oscillatory motions according to the prescription of \citet{GK77} and attains a maximum value of about $10^{13}$ cm$^2$/s near the stellar centre. The turbulent viscous terms in the convective regions tend to damp the induced very short wavelength oscillations \citep{SP97}.

\subsection{Boundary Conditions \label{BCs}}
We apply the usual boundary conditions e.g. \citet{Unno89}  at the stellar centre. At the stellar surface we apply Stefan-Boltzmann's law
 $\frac{\delta F_r}{F} = 4 \, \frac{\delta T}{T} $ and put $\frac{\delta P}{P} = 0 $.
 At the stellar equator we apply the adopted (anti)symmetry of (odd) even modes, at the rotation axis we use the expansion corresponding with the adopted  spherical harmonics (or their derivative) used in the forcing.

\subsection{Numerical Method}
\label{NumMeth}
Equations (\ref{EQ1})-(\ref{EQ6}) are expressed as finite difference equations on a $(r, \theta)$ grid
with 1512 radial grid points and 130 grid points in the interval $\theta = 0 \rightarrow \frac{1}{2} \,\pi$. The solution in the southern hemisphere follows from the adopted parity of the studied oscillation mode. In the radial direction the  MESA code's  staggered mesh is adopted: two cell boundaries where we define $\vec{\xi}$ and $F'_r$, while $P'$ and $T'$ are defined at the cell centre in between. In the numerical procedure three grid levels (both for cell centres and boundaries) are used to allow treatment of second order derivatives in the viscosity terms and the energy diffusion equation.  In the $\theta$ direction we use an equidistant mesh, again with three levels per cell to allow the treatment of second order derivatives. 

The equations of motion and (radial) flux equation are evaluated at the cell boundaries, while the equations of continuity and energy are evaluated at cell centres. 
Starting (at radial level 2) and applying  the 6 or 7 (relaxation version of code) difference equations by moving from the rotation axis towards the equator and using the boundary conditions at the centre all variables at radial level 1, (i.e. $r=0$) can be eliminated. 
By matrix inversions one can set up relations between the unknown perturbation variables at radial level 2 and 3 until the stellar equator is reached. By finally applying the known parity of the oscillation to eliminate the unknown perturbations at the gridpoint beyond the equator it is possible to express each perturbation variable at radial level 2 in terms of (all) the variables at radial level 3. One can now repeat the same procedure from rotation axis to stellar equator at one higher radial level and so on until the surface is reached. After applying the surface boundary conditions for the last stride from rotation axis to equator we obtain the solution at the equator for $r={\rm R_s}$. By backsubstitution in the stored recurrence relations one can work back in reverse order to find the perturbations throughout the 2D stellar oscillation model. 

For a measure of the oscillatory response to the forcing the pressure perturbation is integrated over the upper hemisphere of the star:
\[ \mathcal{R(\overline{\sigma})}=\frac{4 \, \pi}{\rm{M_s}} \, \int^{\rm R_{s}}_{\rm R_{cc}} \int^{\pi/2}_0 \, r^2\, \rho(r) \,\left(\frac{P'(r,\theta)}{P(r)}\right) \,  {\rm P}^m_l(\theta) \,sin{\theta} \, {\rm d} \theta \, {\rm d}r  \]
with $\rm{R_{cc}}$ the convective core boundary; $l$ and $m$ correspond to the applied forcing. By stepping through frequency space and each time recalculating $|\mathcal{R}|$ until a maximum response is passed we search for resonances. We then zoom in onto the resonance and use the resonant solution as input for the relaxation code. 
For this we use the same oscillation equations but without forcing terms and consider the complex oscillation frequency $\overline{\sigma}$ as an extra unknown, thereby introducing non-linear $\overline{\sigma}$ terms. In most cases it requires about ten iterations to converge to the free oscillation mode. The sign of the imaginary part of $\overline{\sigma}$ determines whether the found mode is stable $({\rm Im}(\overline{\sigma})>0)$ or not $({\rm Im}(\overline{\sigma})<0)$. The code was written in Fortran 90 and parallelized to work on multicore machines. On a 16 core node of SARA's national compute cluster LISA (Amsterdam) the code takes about 3 (4) minutes to calculate one iteration for the forced (free) version of the code.

\subsection{Visibility ($\mathcal{V}$) of a mode}
\label{visb}
Since the observations with CoRoT and Kepler deliver detailed photometric information of pulsating stars it is desirable to have a measure of the brightness of the unstable oscillation modes found with our new code. The actual amplitude of the calculated modes can not be determined without greatly increasing the scope and computer requirements of the study by introducing nonlinear effects and coupling between oscillation modes. Because this is beyond the scope of the current study we have to use a simpler way to compare with the observations. There is another important factor that determines the photometric variability: the contribution by the various radiating oscillating surface regions partially cancel, depending on the type of mode and this visibility effect can be easily estimated and used to discriminate between different modes and compare observations with the calculated modes. Recently  \citet{Reese13} have also defined and calculated a (more refined) visibility of modes. They have taken the stellar flattening into account, but their mode calculations are adiabatic. We use the diffusion  equation (\ref{eqf}) for radiative energy transport (or mixing length theory for convective energy transport) for a fully non-adiabatic treatment up to the stellar `surface' and apply (section~\ref{BCs}) Stefan-Boltzmann's law $\frac{\delta F_r}{F}=4 \frac{\delta T}{T}$ at the outer boundary, neglecting details of the stellar atmosphere. 

First the (complex) perturbed radial flux is normalized for the particular mode by dividing it by the maximum modulus attained at some $\theta$-value.
Suppose the observer's inclination angle is $i$ and its azimuthal position angle is given by $\varphi_{obs}$=0. At a point $(\theta, \varphi)$ on the stellar surface the time-dependent real value of the Lagrangian perturbation of the radial surface flux can be expressed as  
\[ \frac{\delta F_r\left(R_{\rm{s}},\theta\right)}{F\left(R_{\rm{s}}\right)} = c_{1} \left(\theta,\varphi\right) \,\cos{(\sigma_r \,t)} + c_{2} \, \left(\theta,\varphi\right)\, \sin{(\sigma_r \,t)} \]
with $\sigma_r = \rm{Re} (\sigma)$ is the real part of the oscillation frequency and 
\[ c_{1} \left(\theta,\varphi\right)= \rm{Re} \left(\frac{\delta F_{r}\left(R_{\rm{s}},\theta\right)}{F\left(R_{\rm{s}}\right)}\right)\, \cos{(m \varphi)} + \rm{Im} \left(\frac{\delta F_{r}\left(R_{\rm{s}},\theta\right)}{F\left(R_{\rm{s}}\right)}\right) \, \sin{(m \varphi)}  \,   \]
\[ c_{2} \left(\theta,\varphi\right)= \rm{Re} \left(\frac{\delta F_{r}\left(R_{\rm{s}},\theta\right)}{F\left(R_{\rm{s}}\right)}\right)\, \sin{(m \varphi)} - \rm{Im} \left(\frac{\delta F_{r}\left(R_{\rm{s}},\theta\right)}{F\left(R_{\rm{s}}\right)}\right) \, \cos{(m \varphi)}  \]

Integrating (numerically) over the area of the stellar hemisphere visible for the observer and dividing by the effective area of the stellar disc we obtain
\[ \mathcal{I}=  \frac{1}{\pi}  \int \int \left[c_{1} \cos{\left(\sigma_r t\right)}+ c_{2} \sin{\left(\sigma_r t \right)} \right]  \: \cos{\gamma} \: \:  \sin{\theta}\, {\rm d} \theta \, {\rm d} \varphi \]
with $ \cos{\gamma}= \left[ \cos{i} \, \cos{\theta} + \sin{i} \, \sin{\theta} \, \cos{\varphi} \right]$
the projection factor. We rewrite this integral as 
\[ \mathcal{I}= C_1 \cos{\left(\sigma_r \, t \right)} + C_2 \sin{\left( \sigma_r \, t \right)} = \mathcal{V} \, \cos{\left(\sigma_r \, t + \alpha \right)} \]
where $\alpha$ is a phase factor. The visibility of the mode is defined as the oscillation amplitude (by normalization a number between 0 and 1)
\begin{equation} 
\mathcal{V} = \sqrt{C_1^2 + C_2^2} 
\end{equation} 

In this paper we  adopt an inclination $i=25$ degrees for the observer, as estimated for the star HD 43317 by \citet{Papi12}. The largest visibility (0.54) occurs for the odd $m=0$ g-mode with 19 radial nodes and frequeny 1.09 cpd. The radial nodes are counted for a fixed value of $\theta$.

\section{Results}
Stability calculations were performed for oscillation modes with $m$-values of 0, 1 and 2 focussed on the frequency range  $<$ 4.5 cpd in the inertial frame where the brightest lines occur. Table~\ref{cpdranges} shows the frequency ranges  where unstable modes are found in both the stellar and inertial frame. From now on all frequencies $\overline{\sigma}$ in the corotating frame are listed in units of the critical rotation frequency $\Omega_{\rm c}=\sqrt{G M^2_{\rm{s}}/R^3_{\rm{s}}}$. Frequencies listed as cycles per day (cpd) are always in the inertial frame. All plots correspond to free oscillation modes, unless stated otherwise (Fig.~\ref{100+3783}).

\subsection{Used coding \mbox{$(l,\tau,m)$} for the oscillation modes \label{mcodes}}
As indicated above we apply forcing terms defined by spherical harmonic indices $(l, m)$ in the oscillation equations and search for resonances with modes compatible with the thus imposed behaviour near the rotation axis and symmetry about the stellar equator.
For a given $m$-value and adopted parity at the equator we choose for the applied forcing the lowest compatible $l$-value, for example for $m=1$ we choose either $l=1$ (even modes) or $l=2$ (odd modes). The oscillation of an odd $m$=1 mode includes components $l=2, 4, 6$  etc. 
In the following modes will be described by three numbers given in parentheses, like (202) or (211) where the first number denotes the $l$-value and the third the $m$-value of the initially applied forcing. The middle number ($\tau$) is either 0 (`spheroidal' forcing) or 1 (`toroidal' forcing). The (211) mode is thus an odd r-mode with $m$=1. We use the same coding for the {\it free} oscillation modes as they have the same symmetries as the corresponding forced solution from which they were derived.
The perturbations, like $P'/P$ or the components of the displacement vector $\vec{\xi}$, are complex quantities as they describe the amplitude and phase of the oscillation. 
\begin{table}
\center
\caption{Frequency ranges where unstable modes are found: respectively in stellar frame (in units of $\Omega_c$) and after `;' in inertial frame (in cycles per day). We have searched no further than $\simeq$ 4.5 cpd in the inertial frame.  Positive or negative frequency values correspond to prograde or retrogade wave motion in stellar/inertial frame. The mode coding (100) etc. is defined in section~\ref{mcodes}. }
\[ 
\begin{array}{|c|c|cc|}  
\hline                                                                              
(l\, \tau \,m) & \overline{\sigma}/\Omega_c>0 \, ; \,\, \mbox{cpd} &  \overline{\sigma}/\Omega_c <0 \, ; \,\, \mbox{cpd}  &  \\
\hline                                                                                             
\hline 
(100) & (1.58 : 0.34) ; (4.32:0.93) & ... & \\
(2 0 0) & (1.69 : 0.33) ; (3.63 : 1.42) & ...  &  \\ 
\hline
(1 0 1) & (0.61 : 0.52) ; (3.02 : 2.79) & (-1.26 : -0.55) ; (-2.08 : -0.14) & \\
(2 0 1) & (1.12 ; 0.38) ; (4.43 : 2.40) & (-1.19 : -0.62) ; (-1.89 ; -0.34) & \\
(2 1 1) & ... & (-0.3273) ;  (0.467) & \\
\hline
(2 0 2) & (0.40 : 0.33) ; (3.81 : 3.63) & (-1.31 : -0.51) ; (-1.54 : 1.32) & \\
(3 0 2) & (0.58 : 0.43) ; (4.31 : 3.89) &  (-1.00 -0.69) ; (-0.01 : 0.83) & \\
(3 1 2) & ... & (-0.28 : -0.25) ; (1.97 : 2.03) & \\   
\hline            
\end{array}
\] 
\label{cpdranges}
\end{table}

 Our simple coding ($l$, $\tau$, $m$) for the calculated modes indicates the basic symmetries, but this gives no information about the actual harmonic content of the mode. 
The Coriolis force generates components with higher $l$-values, see Fig.~\ref{200+6272} and modifies the gravity modes which become `gravito-inertial modes' and contain in principle an infinite number of spherical harmonic components $l$ for a given $m$. In practice their number is evidently limited by the grid resolution.
For $|\overline{\sigma}| > 0.5$ the consecutive unstable g-modes with given $m$-value are often of different harmonic content in the sense that the $l$-value of the dominant (mass average over star) Fourier-Legendre coefficient $(C_l)_{max}$ fluctuates (and the number of radial nodes jumps about)  between consecutive modes. However, for  about $|\overline{\sigma}| < 0.5$ the high radial order components are heavily damped and consecutive unstable modes  are usually in monotonic radial order with a fixed dominant $l$-value and produce the striking sequences of (100) and (200) g-modes in Fig.~\ref{visbl25}.

In the figures below showing components $\xi_{\theta}$ and $\xi_{\varphi}$ of the displacement vector we plot the modulus of these complex quantities.

\subsection{Gravito-inertial modes}
For frequencies in the inertial range $| \overline{\sigma} | < 2\, \Omega_{\rm s}$ gravity waves are substantially modified by rotation and sometimes called gravito-inertial waves. These waves can only propagate between the critical (co)latitudes $\theta_c$ and $\pi - \theta_c$, i.e. in the `equatorial belt' of the rotating star. The critical (co)latitude $\theta_c$ is defined by $\cos{\left(\theta_c\right)}=  \overline{\sigma} / (2 \mbox{$\Omega_{\rm s}$})$. In the sections below  we will simply speak of g-modes when we mean these rotationally modified gravity modes.

Gravito-inertial waves were studied by \citet{SPA95} and \citet{PS97} in tidal forcing calculations of a massive ZAMS star. \citet{Dint99} studied gravito-inertial waves for a stably stratified spherical shell in the Boussinesq approximation. In the latter study these waves were shown to have attractors with focussing on the critical surfaces. However, a small diffusion as found in a normal star cancels their focussing power.  Several new studies of rotational (inertial) waves in fully convective or barotropic stars/planets have appeared, see \citet{PapIv10}, \citet{IvPap10} and \citet{Rieu10}.

\citet{Nein12} claim that gravito-inertial modes are excited by stochastic effects in the hot Be-star HD 51452 as the $\kappa$-mechanism is not effective in such a hot star. They suggest that the gravito-inertial modes found by \citet{Papi12} in HD43317 are also of stochastic nature. However, our current results show that in HD43317 these g-modes can be driven by the $\kappa$-mechanism.

\subsection{r-modes}
The Coriolis force also enables the occurrence of r-modes, first studied as purely toroidal modes in stars by \citet{PP78}.  Using the `traditional approximation' it was discovered by \citet{Sav05} that a class of r-modes (`quasi g-modes') show sufficient density and temperature variations to be destabilized by the $\kappa$-mechanism. \citet{Town05} came almost simultaneously with a similar analysis to the same conclusion, while \citet{Lee06} investigated the stability of these `buoyant' r-modes in more detail with the method of spherical harmonic expansions. We do find a sequence of unstable buoyant r-modes in our calculations, in perfect radial order.

\subsection{Short wavelengths in the stellar interior}
\label{shortwav} 
\begin{figure}[htbp]
{\resizebox{0.5\textwidth} {!}{\includegraphics{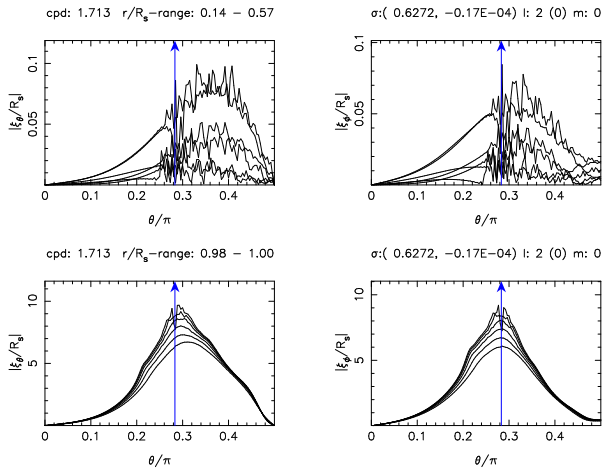}}}
\caption{Modulus of the horizontal components $\xi_{\theta}$ (left panels) and $\xi_{\varphi}$ (right panels) of the displacement vector versus $\theta/\pi$ at some radial meshpoints in the intermediate region outside the convective core (upper panels) and near the stellar surface, including the convective shell, (lower two panels) for a (200) g-mode with frequency 1.713 cpd in the inertial frame. The arrow indicates the critical (co)latitude $\theta_c$.}
      \label{200+6272}
\end{figure}

\begin{figure}[htbp]
{\resizebox{0.5\textwidth} {!}{\includegraphics{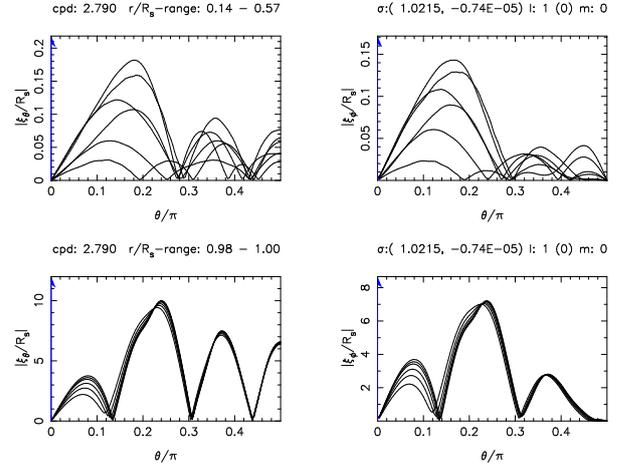}}}
\caption{Modulus of the horizontal components ($\xi_{\theta}$ and $\xi_{\varphi}$) of the displacement vector versus $\theta/\pi$ at some radial meshpoints in the intermediate region outside the convective core (upper panels) and near the stellar surface (lower two panels) for a (100) g-mode with frequency $\overline{\sigma}_r= 1.0215$ (outside the inertial range) in the stellar frame. }
      \label{100+1p022}
\end{figure}

As noted by \citet{PS97} very short wavelength response is excited in the convective core by tidal forcing with frequencies $|\overline{\sigma}| < 2 \Omega_c$ (`inertial range'). These inertial waves can leak out of the core and excite short wavelength gravity waves.  It is interesting that also in the free oscillation solutions the Coriolis force induces short wavelength oscillations near the boundary of the convective core and in the intermediate regions outside the convective core. We find that these  waves can propagate quite far out into the star, see Fig.~\ref{200+6272}. 
One may wonder whether the forced solution that is used as input model is  the cause of the short wavelengths in the free oscillation  solution. To check this we applied a high viscosity $\zeta = 1.0 \times 10^{20}$ cm$^2$/s throughout the star during the first iteration in which the complex frequency $\overline{\sigma}$ is kept fixed. The high viscosity makes the first iterated solution smooth like the disturbances in Fig.~\ref{100+1p022}. But the converged free solution appears independent of this smoothing, again  with a significant short wavelength component in the stellar interior.  

The short wavelength oscillations are limited to the g-mode propagation cavity, i.e. the equatorial belt  between the two critical latitudes.  Further out for $r/R_s > 0.5$ they become weaker and finally disappear near  $r/R_s \simeq 0.7$. This is far below the driving region for the $\kappa$-mechanism which is located near $r/R_s \simeq 0.96$ where $T \simeq 2 \times 10^5$ K. 

For frequencies just outside the inertial range, where the Coriolis force is still significant, it generates higher order spherical harmonic components  but the short wavelengths are completely absent in the star, see Fig.~\ref{100+1p022}.

\begin{figure}[htbp]
{\resizebox{0.5\textwidth} {!}{\includegraphics{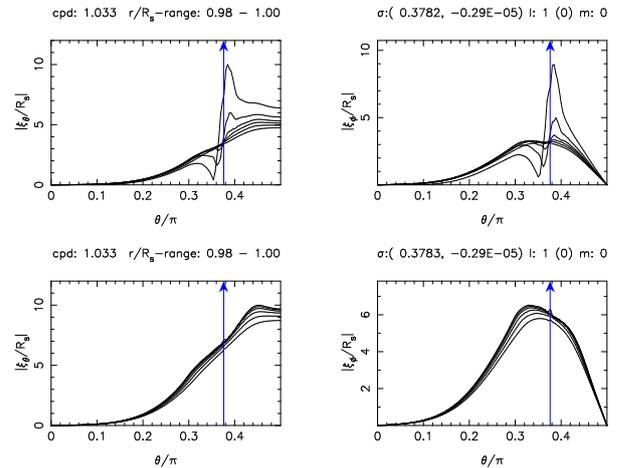}}}
\caption{Modulus of the horizontal components ($\xi_{\theta}$ and $\xi_{\varphi}$) of the displacement vector versus $\theta/\pi$ at some radial meshpoints in the stellar surface region, including the convective shell,  for the (100) g-mode with $\overline{\sigma}=(0.3783, -0.29\times 10^{-5})$ corresponding to 1.033 cpd in the inertial frame. The upper two panels depict the forced solution (with a jump), the lower panels the corresponding free solution with a short wavelength disturbance near $\theta_c$ and a tiny spike. The arrow indicates the critical (co)latitude $\theta_c$.}
      \label{100+3783}
\end{figure}

\subsection{Jump in $\xi_h$ at critical latitude in surface layers (with forcing)}
\label{Disc}
For all studied oscillation modes with $|\overline{\sigma}| \leq 2 \, \mbox{$\Omega_{\rm s}$}$ the solutions with forcing show in the surface region an almost discontinuous jump in the horizontal components of the displacement vector at the critical latitude.  For `buoyant' r-modes this jump occurs at the outer boundary of the convective surface shell, while for g-modes the jump is often generated at the shell's inner boundary. Fig.~\ref{100+3783} shows the horizontal components of the displacements vector in the surface region of a (100) g-mode for both the forced and free solution.  Apparently the discontinuous behaviour at the critical latitudes is generated by the applied forcing, it is absent in the free solutions.  \citet{Terq98}, in their study of the $l$=$m$=2 tides in a non-rotating solar type star, explained the found horizontal displacement $\xi_h$'s  tendency for discontinuous behaviour  near a convective boundary as the consequence of the fluid becoming locally barotropic when the  Brunt-V\"{a}is\"{a}l\"{a} frequency $|N^2| = 0$. In the current calculations with rotation the discontinuity (always focussed at the critical latitude) is also generated at a convective boundary (of the surface shell) and propagates towards the stellar surface.   In the free solutions the jump in $\xi_h$ in the surface layers is absent,  but for g-modes $\xi_h$ does often show in the layers beyond the convective shell small amplitude short wavelength disturbances near the critical latitude $\theta_c$ and a spike or dip at $\theta_c$.

\subsection{Visibility of unstable modes}
As noted above, the best one can do with linear stability calculations is to compare the mode's `visibility' with observed photometric amplitudes.  A first study of non-linear mode coupling in rotating B-stars has been made by \citet{Lee12} applying selection rules for three mode coupling derived in a study by \citet{Schenk02}. Non-linear interaction between an unstable mode and two stable daughter modes will lower the unstable mode's amplitude and can cause the excitation of the two (linearly) stable modes (processes that are ignored in our calculations). Further work is required to understand and apply non-linear interactions between modes in a rotating star.

\begin{figure*}[htbp]
{\resizebox{1.0\textwidth} {!} {\includegraphics{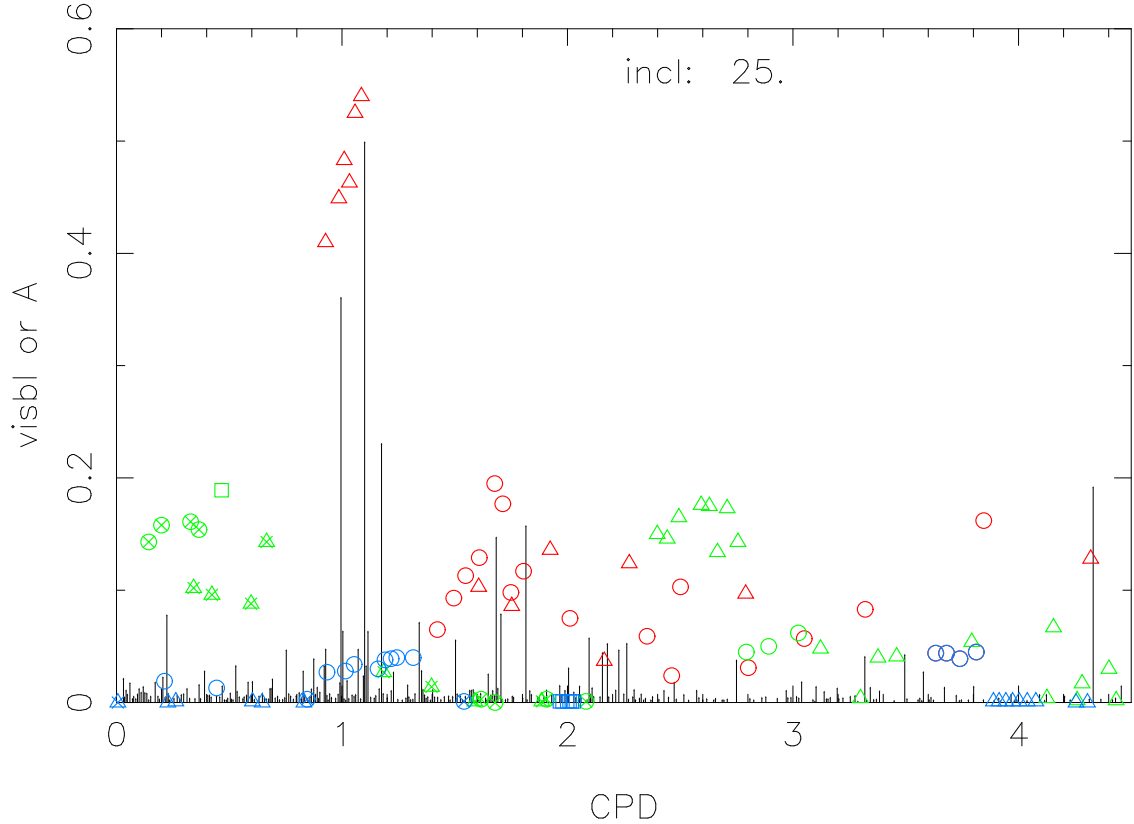}}}
  \caption{Visibility of all overstable modes vs frequency in inertial frame (cpd) projected on P\'{a}pics list of observed photometric amplitudes A (vertical black lines) vs frequency. Adopted inclination is 25 degrees. The photometric amplitudes are here normalized to a maximum of 0.5. Triangles correspond to odd-, circles to even g-modes and squares to (odd) r-modes. A cross inside a symbol means the g-mode is retrograde in the inertial frame. Colour definition:  red $m$=0, green $m$=1 and blue $m$=2.}
  \label{visbl25}
\end{figure*}

 In the current work we can do no better than compare the observed light variations with the visibility of the found unstable modes and see whether on this basis one can identify (some of) the observed photometric variations.
In Fig.~\ref{visbl25} we have plotted the photometric amplitudes (black lines) versus frequency in the inertial frame, as determined by \citet{Papi12} and superposed various symbols representing the here calculated visibility of unstable modes. As expected, the unstable modes with the lowest m-values ($m$=0 and $m$=1) have the highest visibility.
 For the adopted observer's inclination angle of 25 degrees the parity  at the equator of a $m$=0 or 1 mode makes hardly any difference in its visibility. But in Fig.~\ref{visbl25} it can be seen that $m$=2 even  g-modes do have larger visibility than the odd g-modes. Modes with $m \ge 3$ have negligible visibility
 and are ignored.
 \begin{figure}[htbp]
{\resizebox{0.5\textwidth} {!}{\includegraphics{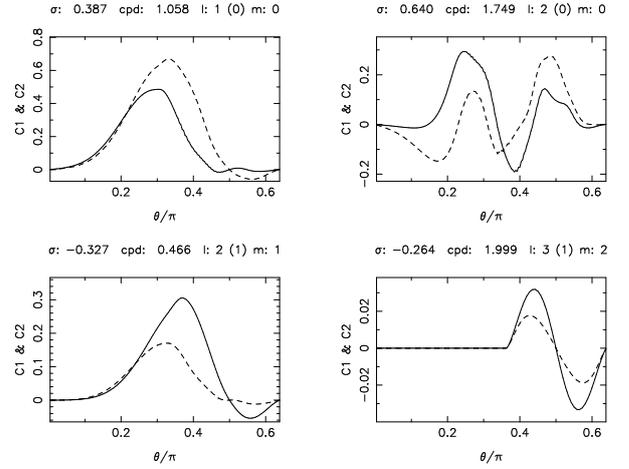}}}
\caption{Visibility integrals $\int c_k(\theta,\varphi)\, \cos(\gamma) \sin(\theta) \,\rm{d} \varphi$ versus $\theta/\pi$. The functions $c_k$ with $k$=1,2 are defined in section~\ref{visb}. The considered ranges of $\varphi$ and $\theta$ correspond to the hemisphere visible for the observer at i=25 degrees. Shown are the results for a (100) and (200) axisymmetric g-mode in the upper panels and a (211) and (312) r-mode in the lower panels.}
      \label{100200c1c2}
\end{figure}
 Fig.~\ref{100200c1c2} shows that the unstable odd axisymmetric (100) g-modes suffer almost no cancellation effect over the $\theta$-range resulting in a high visibility. 
 
The unstable even (200) axisymmetric g-modes have three $\theta$-nodes in the  $\theta$-range 0 - $\pi/2$ and the consequent cancellation effects (Fig.~\ref{100200c1c2}) result in lower visibility. They cover a few observed moderately bright lines. The remaining (100) g-modes with smaller visibility also suffer cancellation as a consequence of two or three nodes in $\theta$. The $m$=1 g-modes with visibilities in the range 0.1-0.2 miss a clear link with observed counterparts, while the even $m$=2 g-modes could correspond with observed weaker `lines' in Fig.~\ref{visbl25}.
We find only one unstable (211) r-mode at 0.47 cpd with moderate visibility (0.19) and a sequence of five unstable (312) r-modes near 2 cpd with negligible visibility. The (312) r-modes suffer both from strong cancellation in the $\varphi$ integration and from cancellation in $\theta$ due to the odd parity at the equator, see Fig.~\ref{100200c1c2}.

 It is consolidating that the found unstable modes with the highest visibility lie close to the frequencies at which the largest photometric amplitudes are observed and correspond even to the observed single moderately strong line located further out at 4.33 cpd.

\subsection{Constant period spacings}
\citet{Papi12} have searched for (nearly) constant period spacings in the light curve of HD 43317 and found a nearly equidistant series of ten peaks with an average period spacing of $\Delta P$= 0.07337 day and another series of seven components with average spacing $\Delta P$ = 0.07385 day for which they claim the semi-constant spacing  is not due to chance. Similar spacings were found for a slow rotator HD 50230 \citep{Groot10}.
We determined the period spacings of modes of a given value of $m$ and do not find nearly constant period spacings of this kind. Most of the period spacings are smaller than at least a factor 4-5 and mostly quite irregular. The current calculations have thus no explanation for the inferred occurrence of (almost) constant period spacings.

\section{Conclusions}
We have performed calculations of a rapidly rotating B-star in which the Coriolis force is dominant or at least substantial with a new method to find unstable oscillation modes.  We have searched for unstable modes with azimuthal index $m$=0, 1 and 2 and defined the visibility of modes by estimating the cancellation effect of the different (perturbed) radiating parts at the stellar surface as seen by the observer. 
By comparing the observed photometric variability (frequencies and brightness) of HD 43317 with the thus determined visibility of modes in Fig.~\ref{visbl25} it is reinforcing that one can discern a global similarity between the two. The most striking is that observationally inferred frequencies corresponding with the largest photometric amplitudes in Fig.~\ref{visbl25} are close to those of the calculated modes with the highest visibility: the odd axisymmetric g-modes. 
Even the bright line at 4.33 cpd appears to almost coincide with an odd axisymmetric mode.
 It seems plausible to identify the observed strongest photometric variations in  HD 43317 as unstable axisymmetric g-modes. 
The connection between the unstable odd $m$=1 g-modes with frequency below 0.8 cpd and between 2 and 3 cpd,  all with moderately high visibility,  and the observed photometric variability is less clear. In most cases no substantial photometric variation is seen for these frequencies. But, as noted above, the neglected non-linear interactions and differential rotation may play a role. 

\begin{acknowledgements} 
The author thanks Bill Paxton for his cooperation in obtaining a smooth composition profile in the MESA results and the referee for critical remarks which have improved the paper. 
\end{acknowledgements}

\bibliographystyle{aa} 
\bibliography{sav.bib} 

\end{document}